\documentclass[fleqn,letter]{aa}

\usepackage{amsmath,amssymb}
\usepackage[varg]{txfonts}
\usepackage{natbib}
\bibpunct{(}{)}{;}{a}{}{,} 
\usepackage[colorlinks=True,
            urlcolor=blue,
            linkcolor=blue,
            citecolor= blue,
            dvips]{hyperref}
\usepackage{booktabs}
\usepackage{longtable}
\usepackage[dvips]{graphicx}

\newcommand{\avg}[1]{\left\langle #1 \right\rangle}
\def\BI{\hbox{$\avg{\rm B_{\rm I}}$}}
\def\BV{\hbox{$\avg{\rm B_{\rm V}}$}}
\def\kG{\hbox{$\rm kG$}}
\def\Prot{\hbox{$P_{\rm rot}$}}
\def\vsini{\hbox{$v\sin i$}}
\def\enumi{\hbox{\itshape (i)}}
\def\enumii{\hbox{\itshape (ii)}}
\def\enumiii{\hbox{\itshape (iii)}}

\begin{document}

\title{What controls the magnetic geometry of M dwarfs?}

\author{T.~Gastine\inst{\ref{mps}} \and J.~Morin\inst{\ref{astro}} \and 
L.~Duarte\inst{\ref{mps}} \and A.~Reiners\inst{\ref{astro}} \and
U.~R.~Christensen\inst{\ref{mps}} \and J.~Wicht\inst{\ref{mps}}}
\institute{Max Planck Institut f\"ur Sonnensystemforschung, 
Max-Planck-Stra{\ss}e 2, 37191 Katlenburg Lindau, Germany \\
\email{gastine@mps.mpg.de} \label{mps}\and Institut f\"ur Astrophysik, 
Georg-August-Universit\"at G\"ottingen, Friedrich-Hund Platz,
37077 G\"ottingen, Germany \label{astro}}


\date{\today, $Revision: 1.107 $}

\abstract
{Observations of rapidly rotating M dwarfs show a
broad variety of large-scale magnetic fields encompassing
dipole-dominated and multipolar geometries.
In dynamo models, the relative importance of inertia in the force balance
-- quantified by the local Rossby number -- is known to have a strong impact on
the magnetic field geometry.}
{We aim to assess the relevance of the local Rossby number in controlling the
large-scale
magnetic field geometry of M dwarfs.}
{We explore the similarities between anelastic dynamo models in
spherical shells and observations of active M-dwarfs, focusing on field
geometries derived from spectropolarimetric studies. To do so, we construct
observation-based quantities aimed to reflect the diagnostic parameters employed
in numerical models.} 
{The transition between dipole-dominated and multipolar
large-scale fields in early to mid M dwarfs is tentatively attributed to
a Rossby number threshold. We interpret late M dwarfs magnetism to result
from a dynamo bistability occurring at low Rossby number. By analogy
with numerical models, we expect different amplitudes of differential rotation
on the two dynamo branches.}
{}


\maketitle

\section{Introduction}

M dwarfs -- the lowest-mass stars of the main sequence -- are of prime interest
to study stellar dynamos operating in physical conditions quite remote from the
solar case. During the past few years, their surface magnetic fields have been
investigated using two complementary approaches \cite[for recent reviews
see][]{Donati09-araa, Reiners12-lrsp}. On the one hand, with spectroscopy in
unpolarized light (i.e. measuring the Stokes parameter I) the average magnetic field strengths (hereafter 
referred to as \BI) of tens
of stars spanning the whole M-dwarf spectral class
have been derived. From such measurements \cite{Reiners07} have concluded
that no significant change in \BI\ occurs when stars become fully convective
(spectral type M3/M4) and \cite{Reiners09a} could constrain the
rotation-magnetic field relation (\BI\ increases towards short rotation periods
until it reaches a plateau at $\sim 3~\kG$).
On the other hand, the measurement of circular polarization (Stokes~V) in
spectral lines of
M dwarfs provides a constraint on the orientation and polarity of the
field at large and intermediate scales. Modelling of the surface field
performed with Zeeman-Doppler imaging \cite[ZDI, ][]{Semel89} on a sample of
about 20 M0-M8 dwarfs points towards a broad variety of magnetic field
geometries. Partly-convective stars as well as a few fully-convective ones
feature complex magnetic structures \citep[e.g.][]{Donati08b, lateM} while
most fully-convective ones host a strong axial dipole component \cite[][]{midM,
Morin08a}.  

Explaining such a diversity in the magnetic field geometry is one of the main
goals of stellar dynamo theory. However, as the numerical simulations
cannot directly access the parameter regime where such dynamos are thought to
operate, asymptotic scaling laws are of prime interest to check the relevance of
the dynamo models \citep{Christensen10}. Scaling laws derived from
geodynamo-like models (i.e. incompressible flow and no-slip boundaries)
successfully fit the magnetic field strength of a broad range of astrophysical
objects, encompassing Earth, Jupiter and some rapidly-rotating M dwarfs
\citep{Christensen09}. Such similarities between the
dynamos in planets and rapidly-rotating stars motivate the comparison of some
planetary dynamo results with observations of active M dwarfs.

In geodynamo models, the relative contribution of inertia and
Coriolis force in the global balance is known to have a strong impact on the
magnetic
field geometry \citep[e.g.][]{Sreenivasan06}. \cite{Christensen06} suggest that
the ratio of these two forces can be quantified by the so-called
``local Rossby number'' defined by $\text{Ro}_l=u_{\text{rms}}/\Omega l$,
$l$ being the typical flow lengthscale. This dimensionless parameter
has been found to be a rather universal quantity that allows to separate dipolar
and multipolar dynamo models: a sharp transition between these two types of
dynamo occurs around $\text{Ro}_l\simeq 0.1$. However, this
has been recently challenged by some models that employed stress-free
mechanical boundary conditions, more appropriate when modelling stellar dynamos
\citep[e.g.][]{Goudard08}. \cite{Simitev09} for instance found that a
dipolar and a multipolar magnetic field can coexist at
the same parameter regime depending on the initial condition. This magnetic
bistability phenomenon can lead to multipolar solutions even for $\text{Ro}_l
< 0.1$ \citep{Schrinner12}.

In addition, most of these studies have been conducted under the
Boussinesq approximation that assumes the reference state to be constant with
radius. While this assumption is suitable for
modelling the geodynamo, it becomes questionable in the stellar
interiors, where density increases by several orders of magnitude.
Anelastic and compressible dynamo models indeed indicate a strong
influence of the density contrast on the geometry of the magnetic field:
some weakly stratified simulations of \cite{Dobler06} are dipole-dominated,
while the strongly stratified models of \cite{Browning08} are
multipolar. The parameter study of \cite{Gastine12a} extends the
bistability phenomenon to a broader parameter range and confirms
that strongly stratified models tend to produce multipolar dynamos.

Following the conclusions by \cite{Schrinner12} and \cite{Gastine12a}
(hereafter GDW12),
the main aim of this letter is to compare the results of the later study
with spectropolarimetric observations of M dwarfs. 

\section{The dynamo model}
\label{sec:model}

We consider MHD simulations of a conducting anelastic fluid in spherical shells
rotating at a constant rotation rate $\Omega$ about the $z$-axis. Following
\cite{Glatz1}, the governing MHD equations are
non-dimensionalised using the shell thickness $d=r_o-r_i$ as the reference
lengthscale and $\Omega^{-1}$ as the time unit.
Our dynamo model results are then characterised by several dimensionless
diagnostic parameters. The rms flow velocity for instance is given by the Rossby
number $\text{Ro}=u_{\text{rms}}/\Omega d$. Following \cite{Christensen06}, we
also employ the aforementioned local Rossby number $\text{Ro}_l=  \text{Ro}\,
\bar{\ell}_u/\pi $,
that is known to be a more appropriate measure to quantify the impact of inertia
on the magnetic field geometry. The mean spherical harmonic degree
$\bar{\ell}_u$ is obtained from the kinetic energy spectrum and relates to the
typical flow lengthscale $l$ through:

\begin{equation}
 l = \pi\, d/\bar{\ell}_u \quad\text{with}\quad\bar{\ell}_u =
\sum_\ell\ell\dfrac{\langle \vec{u}_\ell \cdot \vec{u}_\ell
\rangle}{\langle \vec{u}\cdot\vec{u}\rangle},
\end{equation}
where $\vec{u}_\ell$ is the flow at a given spherical harmonic degree $\ell$ and
the brackets correspond to an average over time and radius.

The magnetic field strength is measured by the Elsasser number
$\Lambda=B_{\text{rms}}^2 /\rho\mu\lambda\Omega$, where $\rho$ is the
density, and $\mu$ and $\lambda$ are the magnetic
permeability and diffusivity. The
geometry of the surface field is quantified by its dipolarity 
$f_\text{dip}=\vec{B}_{\ell=1,m=0}^{2}(r=r_o)/\sum_{\ell,m}^{\ell_{\text{max}}
}\vec{B}_{\ell, m } ^ { 2 } (r=r_o)$, the ratio of the magnetic energy of the
dipole to the magnetic energy contained in spherical harmonic degrees up to
$\ell_{\text{max}}=11$ \citep[see][]{Christensen06}.

The dimensionless MHD equations are advanced in time with the spectral code
MagIC \citep{Wicht02,Gastine12} that uses the anelastic formulation of
\cite{Lantz99} and has been validated against several dynamo
benchmarks \citep{Jones11}. We rely in the following on the results of the
parameter study of GDW12. Although these simulations have
been initially tailored to model the dynamo of giant planets, the
differences in the reference properties (i.e. density and gravity
profiles) are not expected to be of prime influence. Similarities
between dynamos in planets and low-mass stars are indeed emphasised by
geodynamo-based scaling laws \citep{Christensen10} and previous Boussinesq
models employed in the stellar dynamo context
\citep[e.g.][]{Schrinner12,Simitev12}.

\section{Spectropolarimetric observations}
\label{sec:observations}

Spectropolarimetric observations of 23 active M0-M8 dwarfs with
rotation periods ranging from 0.4 to 19 days have been carried out. For each
star at least one time-series of unpolarized and circularly polarized
spectra sampling a few rotation periods has been obtained. The data reduction
and analysis is detailed by \cite{Donati06, Donati08b} and \cite{midM,
Morin08a,lateM}.

The relative importance of inertia with respect to the Coriolis force in the
convection zone of these stars is assessed through an empirical Rossby number
given by
\begin{equation}
  \text{Ro}_{\text{emp}} = \dfrac{\Prot}{\tau_{\text{conv}}},
  \label{eq:Rossby-obs}
\end{equation}
where $\tau_{\text{conv}}$ is the empirical turnover timescale of convection
based on the rotation-activity relation \citep{Kiraga07}.
This Rossby number misses explicitly the flow lengthscale  $l$
involved in $\text{Ro}_l$.
However, as $\tau_{\text{conv}}$ is based on the average convective turnover
time it encompasses this scale information to some extent. We thus use
$\text{Ro}_{\text{emp}}$ as our best available proxy for $\text{Ro}_l$.

For each obtained spectrum, an average line profile with increased
signal-to-noise ratio is computed using the least-squares deconvolution
technique \cite[LSD, ][]{Donati97b}. Each time-series of LSD profiles is
modelled with ZDI, resulting in a map of the large-scale component of the
surface magnetic field vector that satisfies a maximum-entropy criterion. The
large-scale magnetic fields of most of these stars fall into two distinct
groups: one is dominated by a strong axial dipolar component and the other by a
much weaker and non-axisymmetric field. We could not identify any
ZDI reconstruction bias that would spuriously produce such a
separation, and exclude an effect of limited resolution as both groups
span largely overlapping ranges of \vsini.

Following \cite{Morin11}, we define an Elsasser number based on the
averaged unsigned large-scale magnetic field  \BV\ which roughly characterises
the ratio of Lorentz and Coriolis forces. $\eta$ is obtained
by rescaling a reference solar magnetic diffusivity of $\eta_\odot=10^{11}~{\rm
cm}^2 {\rm s}^{-1}$ \citep{Stix89} to the M dwarfs with a MLT
argument. We also consider the fraction of the magnetic energy that is
recovered in the axial dipole mode in ZDI maps. The spatial resolution of such
maps mostly depends on \vsini~and the actual degree and order $\ell_\text{max}$
up to which the reconstruction can be
performed ranges from 4 to 10, although very little energy is recovered in modes
with $4<\ell\leq10$. We therefore directly compare this quantity to the
dipolarity employed in numerical models and term them both $f_\text{dip}$. We
however note that in simulations, $f_\text{dip}$ does not strongly depend on the
chosen $\ell_\text{max}$, whereas for the observation-based  dipolarity,
considering the ratio of magnetic energy in the axial dipole relative to the
total magnetic energy derived from unpolarized spectroscopy (instead of the
large-scale magnetic energy derived from spectropolarimetric data with ZDI)
would lead to much lower values of $f_\text{dip}$ \cite[cf. ][]{Reiners09b}.
We attribute this difference to the
low magnetic Reynolds number ($\text{Rm}\sim 100-500$) accessible by numerical
simulations which does not allow for a significant small-scale field to be
generated  -- hence the weak dependence of $f_\text{dip}$ on $\ell_\text{max}$
-- while in stellar interiors large-scale and small-scale dynamo action
likely coexist \cite[e.g.][]{CattaneoH09}.

\section{Results}
\label{sec:results}

\begin{figure}
\centering
\includegraphics[width=9cm]{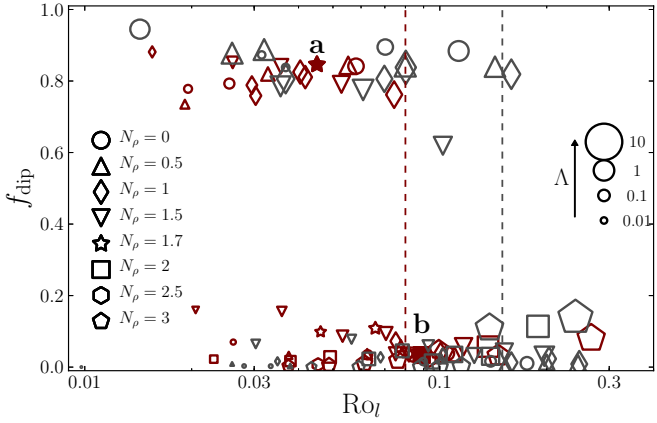}
\caption{\small $f_\text{dip}$ plotted against $\text{Ro}_l$ in
anelastic dynamo models. Red  (grey) symbols correspond to simulations in thick
(thin) shells ($r_i/r_o=0.2$ and $r_i/r_o=0.6$, respectively)  and their size is
scaled according to the value of the surface field, expressed in units of the
square root of the Elsasser number. Each type of symbols corresponds to a
density contrast. The two closed symbols correspond to two cases further
discussed in Fig.~\ref{fig:BrSim}. The vertical lines mark the
tentative limits for dipolar dynamos.}
\label{fig:dipSim}
\end{figure}

\begin{figure}
\centering
\includegraphics[width=9cm]{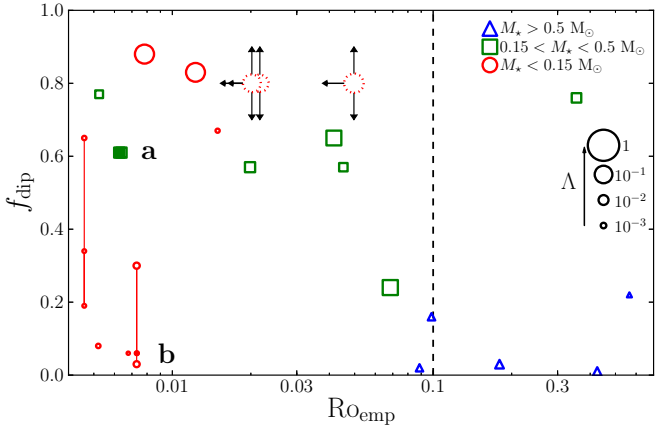}
\caption{\small Observational counterpart of Fig.~\ref{fig:dipSim}. The
vertical dashed line marks the
tentative limit for the dipolar regime. For the two stars
exhibiting the largest temporal variations, the individual epochs are shown and
connected by a vertical red line. Dotted red circles with black arrows
correspond to stars from \cite{lateM} for which a definite ZDI reconstruction
could not be achieved, but an upper limit for the rotation period and an
estimate of dipolarity were derived. The two closed symbols correspond to
two selected stars discussed in Fig.~\ref{fig:BrObs}.}
\label{fig:dipObs}
\end{figure}

Figure~\ref{fig:dipSim} shows $f_\text{dip}$ versus $\text{Ro}_l$ in the
numerical models, which use the anelastic approximation
with density contrasts up to $N_\rho = \ln (\rho_{\text{bot}}/\rho_{\text{top}})
=3$ and stress-free (or mixed) boundary conditions. Figure~\ref{fig:dipObs}
displays the relative dipole strength of M stars against $\text{Ro}_\text{emp}$
derived from spectropolarimetric observations.

The numerical models cluster in two distinct dynamo branches: the upper branch
corresponds to the dipole-dominated regime ($f_\text{dip} > 0.6$), while the
dynamo models belonging to the lower branch
have a multipolar magnetic field ($f_\text{dip} < 0.2$). In agreement
with the geodynamo models, the dipolar branch is bounded by a maximum
$\text{Ro}_l$, beyond which all the models become multipolar
\citep{Christensen06}. Although a small dependence on the shell thickness is
noticeable, the dipolar branch vanishes around $\text{Ro}_l \simeq 0.1$ for
both spherical shell geometries considered by GDW12. However, in
contrast with most of the previous Boussinesq studies, the multipolar branch
also extends well below $\text{Ro}_l \simeq 0.1$, where both dipolar and
multipolar solutions can coexist \citep[see also][]{Schrinner12}. In the
parameter range covered by our numerical models, bistability of the magnetic
field is in fact quite common, meaning that both a dipole-dominated and a
multipolar field are possible configurations at the same set of parameters. The
type of solution is then selected by the geometry and the amplitude of the
initial seed magnetic field \citep{Busse06,Simitev09}. The multipolar
branch found at $\text{Ro}_l<0.1$ is partly composed by the multipolar component
of such bistable dynamos but also encompasses all the stratified models with
$N_\rho\geq2$, for which no dipolar solutions are found (GDW12).

The comparison between these results and the observations of stellar
magnetic fields suffers from the
difficulty to relate the diagnostic parameters of the dynamo models
(i.e. $\text{Ro}_l$, $\Lambda$ and $f_\text{dip}$) to their
stellar counterparts. Within these limits, the separation into two dynamo
branches seems to be relevant to the sample of active M dwarfs displayed in
Fig.~\ref{fig:dipObs}. In fact, all the early M stars (with $M_\star >
0.5\,\text{M}_\odot$) show multipolar magnetic fields with $f_\text{dip} < 0.2$;
being slow rotators they fall into the $\text{Ro}_\text{emp} \geq 0.1$ regime
\citep{Reiners12}. The observations of mid M dwarfs
suggest a possible transition between dipole-dominated and multipolar magnetic
fields close to $\text{Ro}_\text{emp} \sim 0.1$, although CE~Boo does not fit
into this picture (green square in the upper right corner
of Fig.~\ref{fig:dipObs}). Late M dwarfs (with $M_\star <
0.15\,\text{M}_\odot$) seem to operate in two different dynamo regimes: the
first ones show a strong dipolar magnetic field, while others present a weaker
multipolar magnetic structure with a significant time variability (emphasised by
the vertical red lines in Fig.~\ref{fig:dipObs}). These important time
variations of the dipole strength are also frequently observed in multipolar
dynamo models with low $\text{Ro}_l$ \citep[e.g.][GDW12]{Schrinner12} but
are not visible in Fig.~\ref{fig:dipSim}, where time-averaged properties are
considered.  
We note that the values of $\tau_\text{conv}$ entering Ro$_\text{emp}$
are poorly constrained for M dwarfs. Assuming a stronger
(weaker) mass-dependence would mostly expand (shrink) the x-axis of
Fig.~\ref{fig:dipObs} without changing our conclusions. The
uncertainties on $f_\text{dip}$ typically lie in the range
0.1-0.3 \cite[see discussion in ][]{lateM}, which does not question the
identification of two distinct branches.

\begin{figure}
\centering
\includegraphics[width=9cm]{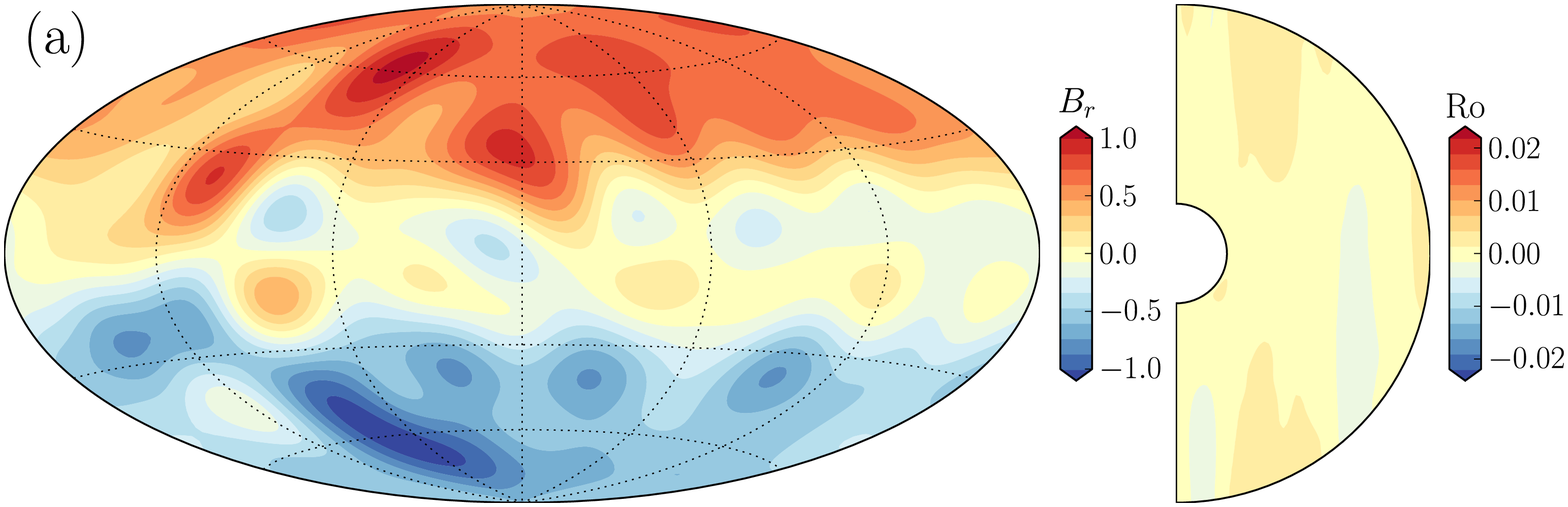}
\includegraphics[width=9cm]{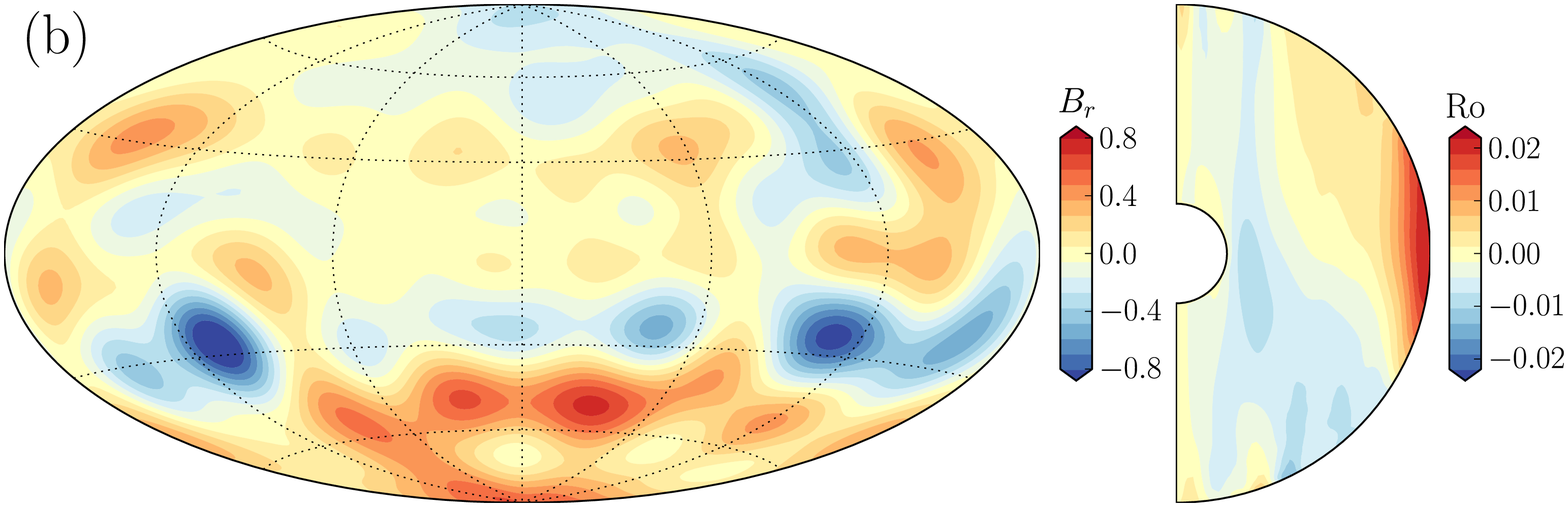}
\caption{\small Surface radial magnetic field
$B_r(r=r_0)$ and axisymmetric zonal flows $\overline{u_\phi}$ for a
dipolar dynamo model with $N_\rho=1.7$ (\textbf{a}) and a multipolar one
with  $N_\rho=2$ (\textbf{b}). The maps of $B_r$ have been low-pass filtered
up to $\ell_{\text{max}} = 10$. Magnetic fields are expressed in
units of the square root of the Elsasser number and velocities  in units  of 
the Rossby number.} 
\label{fig:BrSim}
\end{figure}

\begin{figure}
\centering
\includegraphics[width=9cm]{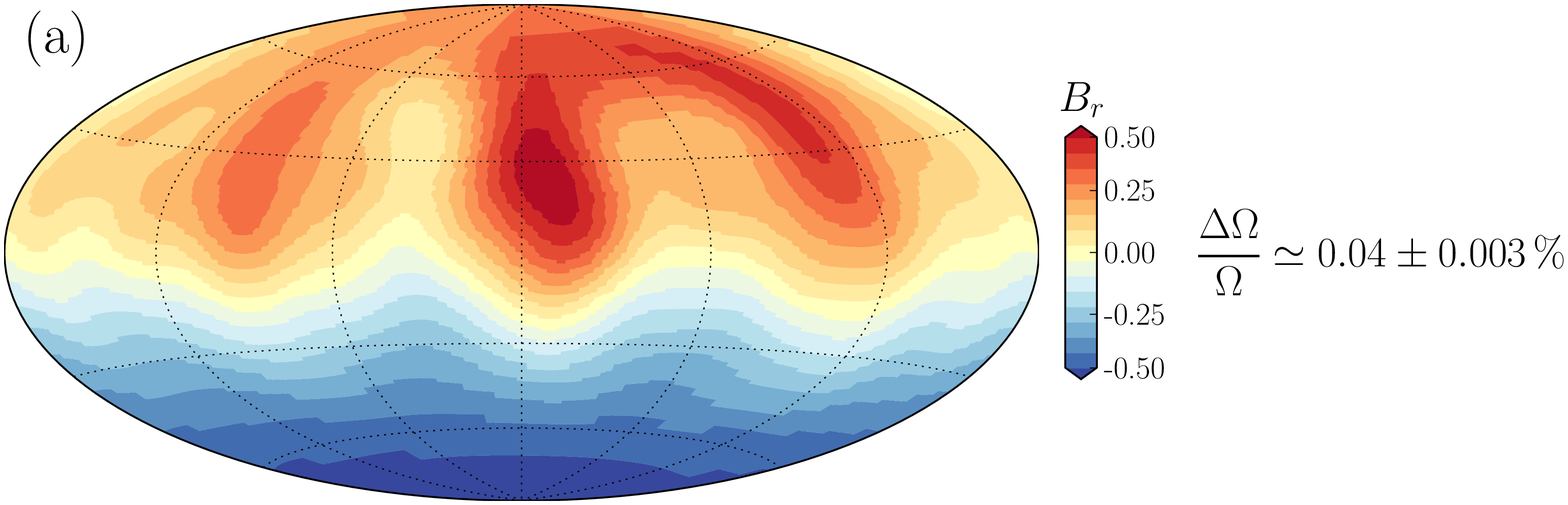}
\includegraphics[width=9cm]{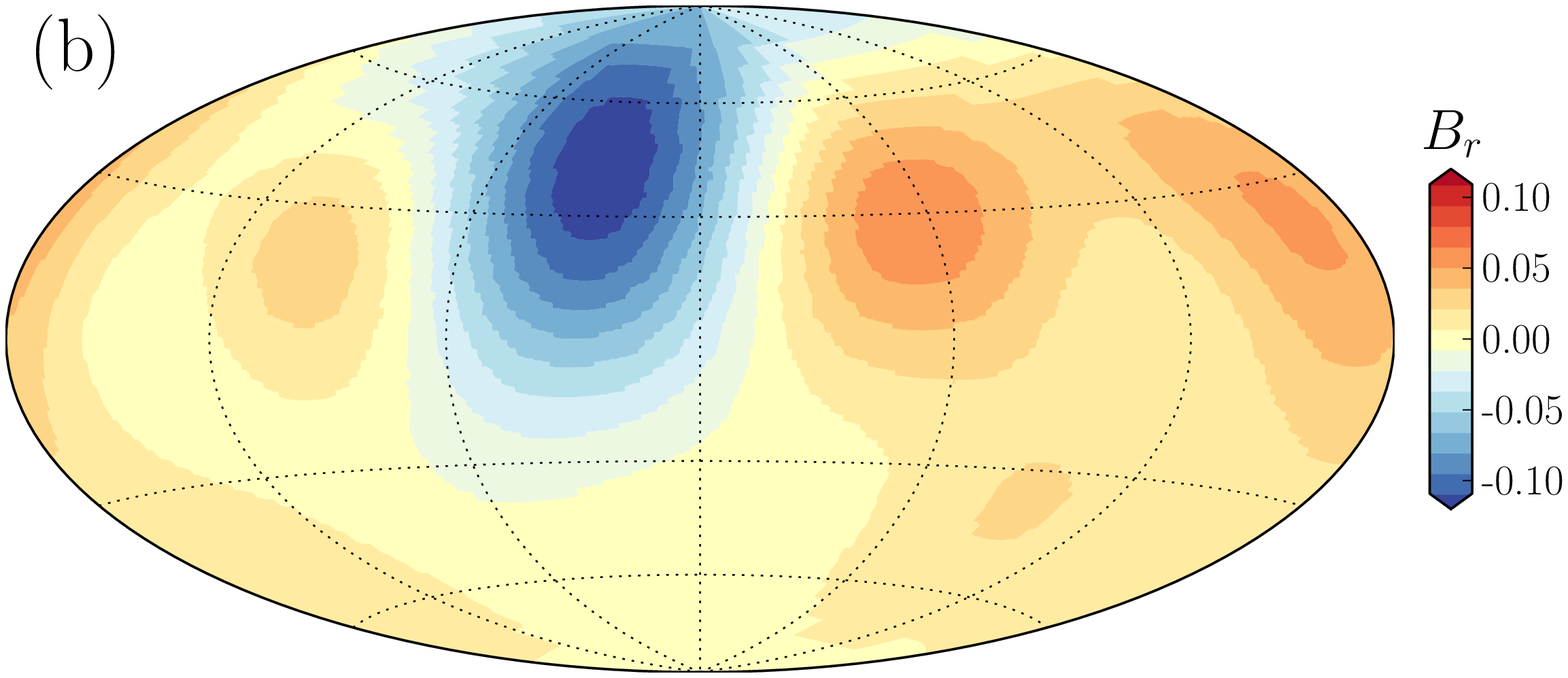}
\caption{\small Surface radial magnetic field of V374~Peg
(\textbf{a}) and GJ~1245~B (\textbf{b}) recovered with ZDI from
spectropolarimetric observations. The field has been
reconstructed up to $\ell_{\text{max}}=10$ (4) for V374~Peg (GJ~1245~B). Surface
differential rotation of V374~Peg has been derived by \cite{Morin08a} from
spectropolarimetric observations, while this was not possible for GJ~1245~B.
Magnetic fields are expressed in units of the square root of the Elsasser
number.} 
\label{fig:BrObs}
\end{figure}

The two dynamo branches found in our numerical models also differ by their main
force balance, at least in the bistability region (i.e. $\text{Ro}_l < 0.1$).
The dipolar branch encompasses models with Elsasser number around unity
that suggest a first-order contribution of the
Lorentz force in the main balance \citep[the dynamo then operates under the
so-called ``magnetostrophic balance'', e.g.][]{Sreenivasan06}. In
contrast, at low $\text{Ro}_l$, the models belonging to the multipolar branch
have weaker magnetic fields (see Fig.~\ref{fig:dipSim} and Tab.~2
in GDW12), meaning that the Lorentz force may not enter in the
first-order balance. In that case, significant geostrophic zonal flows (i.e.
aligned on coaxial cylinders) can possibly develop \citep[e.g.][]{Gastine12}.
Note that beyond $\text{Ro}_l > 0.1$, the multipolar dynamos are strong
enough to yield a magnetostrophic force balance due to larger Rayleigh numbers.
This correspondence between high dipolar fraction and strong large-scale
magnetic field is also present in spectropolarimetric observations
\citep[][]{Morin11}. Figure~\ref{fig:BrSim} illustrates these two types of
dynamos for two typical models with $\text{Ro}_l < 0.1$. While the
upper panel shows a dipole-dominated solution with a very weak differential
rotation, the lower one has a multipolar field that goes along with
significant zonal flows that become nearly geostrophic close to the equator.
These strong zonal winds affect the dynamo mechanism via the
$\Omega$-effect, i.e. the production of toroidal field by the shear.
This $\Omega$-effect plays only a minor role in the field production of the
dipole-dominated models that can be categorised as ``$\alpha^2$-dynamos'',
according to the mean-field description \citep[e.g.][]{Chabrier06}. In
contrast, dynamos on the multipolar branch can be classified as
$\alpha\Omega$ or $\alpha^2\Omega$, at least in the low $\text{Ro}_l$ regime.

To a certain extent, these differences are also noticeable in the reconstructed
fields of M dwarfs as illustrated in
Fig.~\ref{fig:BrObs}. While V374~Peg (upper panel) has a dipole-dominated
magnetic field and a very weak differential rotation, GJ~1245~B (lower panel)
presents a weaker amplitude multipolar field with important time variability
(one of the vertical red lines in Fig.~\ref{fig:dipObs}).
If, as suggested in the present study, the multipolar fields observed in several
late M dwarfs are the consequence of a dynamo bistability occurring at low
$\text{Ro}_l$, the numerical models would then suggest a significant
differential rotation in these stars ($\Delta\Omega/\Omega \sim 5\%$ in the
Fig.~\ref{fig:BrSim}b model). However, this cannot be verified from
the available data as surface differential rotation has been determined for
one star of the sample only (a dipolar one, see Fig.~\ref{fig:BrObs}a).

\section{Conclusion}
\label{sec:conclusion}

Spectropolarimetric observations of active M dwarfs and dynamo
models show a broad variety of magnetic geometries \citep[see][and references
therein]{Gastine12a,lateM}. In both cases, dipolar
and multipolar large-scale magnetic fields are found to coexist at low Rossby
numbers. In the present letter we critically discuss the analogy between these
two results.

We derive observation-based quantities aimed to reflect the diagnostic
parameters employed in the numerical models ($\text{Ro}_l$, $\Lambda$ and
$f_{\text{dip}}$), although these crude proxies are not expected to provide a
direct quantitative match. Within these limits, we draw
an interesting analogy between the observational parameters and their numerical
counterparts: for large values of the Rossby numbers multipolar fields are
found, while below a critical value around $\text{Ro}_l \sim
\text{Ro}_\text{emp} \sim 0.1$, a bistable region exists where both dipolar and
multipolar fields can be generated.

Several limitations must be noted though. \enumi\ The spectropolarimetric
sample is biased as all stars at high (low) $\text{Ro}_\text{emp}$ are partly
(fully) convective.
Thus it is not yet clear if the change in $f_\text{dip}$ observed around
$\text{Ro}_\text{emp}\sim 0.1$ can be
attributed to a threshold in $\text{Ro}_l$ or rather to the drastic changes
in stellar structure occurring at the fully-convective transition.
\enumii\ As the numerical models of GDW12 do not attempt to model
a tachocline, they might miss some important features of early M
dwarfs magnetism. However these issues do not question  the
validity of the agreement between observations and simulations regarding the
existence of a bistable dynamo regime at low $\text{Ro}_l$ for fully-convective
stars. \enumiii\ In numerical models, the dipolar branch only exists for
mild density contrasts ($N_\rho< 2$), much below the stratification of
stellar interiors. Different assumptions from those
used by GDW12 could possibly extend the dipolar regime towards
higher stratifications, for instance by using different
values of Prandtl numbers \citep{Simitev09} or radius-dependent
properties (e.g. thermal and ohmic diffusivities).

However, the analogy between dynamo simulations and
magnetic properties of M dwarfs can be further assessed with more realistic
numerical models, and additional observations as it implies that: \enumi\
stars with multipolar fields can
be found over the whole parameter range where also dipole-dominated
large-scale fields are observed; \enumii\ in the bistable domain, stars on the
multipolar branch have a much stronger surface differential rotation than stars
hosting dipole-dominated large-scale fields.

\begin{acknowledgements}
TG and LD are supported by the Special Priority Program 1488 (PlanetMag)
of the German Science Foundation. JM is funded by a postdoctoral fellowship of
the Alexander von Humboldt foundation. AR acknowledges research funding from DFG
grant RE 1664/9-1.
\end{acknowledgements}

\bibliographystyle{aa}

\begin{thebibliography}{35}
\expandafter\ifx\csname natexlab\endcsname\relax\def\natexlab#1{#1}\fi

\bibitem[{{Browning}(2008)}]{Browning08}
{Browning}, M.~K. 2008, \apj, 676, 1262

\bibitem[{{Busse} \& {Simitev}(2006)}]{Busse06}
{Busse}, F.~H. \& {Simitev}, R.~D. 2006, GAFD, 100, 341

\bibitem[{{Cattaneo} \& {Hughes}(2009)}]{CattaneoH09}
{Cattaneo}, F. \& {Hughes}, D.~W. 2009, \mnras, 395, L48

\bibitem[{{Chabrier} \& {K{\"u}ker}(2006)}]{Chabrier06}
{Chabrier}, G. \& {K{\"u}ker}, M. 2006, \aap, 446, 1027

\bibitem[{{Christensen}(2010)}]{Christensen10}
{Christensen}, U.~R. 2010, \ssr, 152, 565

\bibitem[{{Christensen} \& {Aubert}(2006)}]{Christensen06}
{Christensen}, U.~R. \& {Aubert}, J. 2006, GJI, 166, 97

\bibitem[{{Christensen} {et~al.}(2009){Christensen}, {Holzwarth}, \&
  {Reiners}}]{Christensen09}
{Christensen}, U.~R., {Holzwarth}, V., \& {Reiners}, A. 2009, \nat, 457, 167

\bibitem[{{Dobler} {et~al.}(2006){Dobler}, {Stix}, \& {Brandenburg}}]{Dobler06}
{Dobler}, W., {Stix}, M., \& {Brandenburg}, A. 2006, \apj, 638, 336

\bibitem[{{Donati} {et~al.}(2006){Donati}, {Forveille}, {Cameron}, {Barnes},
  {Delfosse}, {Jardine}, \& {Valenti}}]{Donati06}
{Donati}, J.-F., {Forveille}, T., {Cameron}, A.~C., {et~al.} 2006, Science,
  311, 633

\bibitem[{{Donati} \& {Landstreet}(2009)}]{Donati09-araa}
{Donati}, J.-F. \& {Landstreet}, J.~D. 2009, \araa, 47, 333

\bibitem[{{Donati} {et~al.}(2008){Donati}, {Morin}, {Petit}, {Delfosse},
  {Forveille}, {Auri{\`e}re}, {Cabanac}, {Dintrans}, {Fares}, {Gastine},
  {Jardine}, {Ligni{\`e}res}, {Paletou}, {Velez}, \& {Th{\'e}ado}}]{Donati08b}
{Donati}, J.-F., {Morin}, J., {Petit}, P., {et~al.} 2008, \mnras, 390, 545

\bibitem[{{Donati} {et~al.}(1997){Donati}, {Semel}, {Carter}, {Rees}, {Collier
  Cameron}, \& {Meerschweinchen}}]{Donati97b}
{Donati}, J.-F., {Semel}, M., {Carter}, B.~D., {et~al.} 1997, \mnras, 291, 658

\bibitem[{{Gastine} {et~al.}(2012){Gastine}, {Duarte}, \& {Wicht}}]{Gastine12a}
{Gastine}, T., {Duarte}, L., \& {Wicht}, J. 2012, \aap, 546, A19, [GDW12]

\bibitem[{{Gastine} \& {Wicht}(2012)}]{Gastine12}
{Gastine}, T. \& {Wicht}, J. 2012, \icarus, 219, 428

\bibitem[{{Gilman} \& {Glatzmaier}(1981)}]{Glatz1}
{Gilman}, P.~A. \& {Glatzmaier}, G.~A. 1981, \apjs, 45, 335

\bibitem[{{Goudard} \& {Dormy}(2008)}]{Goudard08}
{Goudard}, L. \& {Dormy}, E. 2008, Europhysics Letters, 83, 59001

\bibitem[{{Jones} {et~al.}(2011){Jones}, {Boronski}, {Brun}, {Glatzmaier},
  {Gastine}, {Miesch}, \& {Wicht}}]{Jones11}
{Jones}, C.~A., {Boronski}, P., {Brun}, A.~S., {et~al.} 2011, \icarus, 216, 120

\bibitem[{{Kiraga} \& {Stepien}(2007)}]{Kiraga07}
{Kiraga}, M. \& {Stepien}, K. 2007, Acta Astronomica, 57, 149

\bibitem[{{Lantz} \& {Fan}(1999)}]{Lantz99}
{Lantz}, S.~R. \& {Fan}, Y. 1999, \apjs, 121, 247

\bibitem[{{Morin} {et~al.}(2008{\natexlab{a}}){Morin}, {Donati}, {Petit},
  {Delfosse}, {Forveille}, {Albert}, {Auri{\`e}re}, {Cabanac}, {Dintrans},
  {Fares}, {Gastine}, {Jardine}, {Ligni{\`e}res}, {Paletou}, {Ramirez Velez},
  \& {Th{\'e}ado}}]{midM}
{Morin}, J., {Donati}, J., {Petit}, P., {et~al.} 2008{\natexlab{a}}, \mnras,
  390, 567

\bibitem[{{Morin} {et~al.}(2008{\natexlab{b}}){Morin}, {Donati}, {Forveille},
  {Delfosse}, {Dobler}, {Petit}, {Jardine}, {Cameron}, {Albert}, {Manset},
  {Dintrans}, {Chabrier}, \& {Valenti}}]{Morin08a}
{Morin}, J., {Donati}, J.-F., {Forveille}, T., {et~al.} 2008{\natexlab{b}},
  \mnras, 384, 77

\bibitem[{{Morin} {et~al.}(2010){Morin}, {Donati}, {Petit}, {Delfosse},
  {Forveille}, \& {Jardine}}]{lateM}
{Morin}, J., {Donati}, J.-F., {Petit}, P., {et~al.} 2010, \mnras, 407, 2269

\bibitem[{{Morin} {et~al.}(2011){Morin}, {Dormy}, {Schrinner}, \&
  {Donati}}]{Morin11}
{Morin}, J., {Dormy}, E., {Schrinner}, M., \& {Donati}, J.-F. 2011, \mnras,
  418, L133

\bibitem[{{Reiners}(2012)}]{Reiners12-lrsp}
{Reiners}, A. 2012, Living Reviews in Solar Physics, 9, 1

\bibitem[{{Reiners} \& {Basri}(2007)}]{Reiners07}
{Reiners}, A. \& {Basri}, G. 2007, \apj, 656, 1121

\bibitem[{{Reiners} \& {Basri}(2009)}]{Reiners09b}
{Reiners}, A. \& {Basri}, G. 2009, \aap, 496, 787

\bibitem[{{Reiners} {et~al.}(2009){Reiners}, {Basri}, \&
  {Browning}}]{Reiners09a}
{Reiners}, A., {Basri}, G., \& {Browning}, M. 2009, \apj, 692, 538

\bibitem[{{Reiners} \& {Mohanty}(2012)}]{Reiners12}
{Reiners}, A. \& {Mohanty}, S. 2012, \apj, 746, 43

\bibitem[{{Schrinner} {et~al.}(2012){Schrinner}, {Petitdemange}, \&
  {Dormy}}]{Schrinner12}
{Schrinner}, M., {Petitdemange}, L., \& {Dormy}, E. 2012, \apj, 752, 121

\bibitem[{{Semel}(1989)}]{Semel89}
{Semel}, M. 1989, \aap, 225, 456

\bibitem[{{Simitev} \& {Busse}(2009)}]{Simitev09}
{Simitev}, R.~D. \& {Busse}, F.~H. 2009, Europhysics Letters, 85, 19001

\bibitem[{{Simitev} \& {Busse}(2012)}]{Simitev12}
{Simitev}, R.~D. \& {Busse}, F.~H. 2012, \apj, 749, 9

\bibitem[{{Sreenivasan} \& {Jones}(2006)}]{Sreenivasan06}
{Sreenivasan}, B. \& {Jones}, C.~A. 2006, GJI, 164, 467

\bibitem[{{Stix}(1989)}]{Stix89}
{Stix}, M. 1989, {The Sun. an Introduction} (Berlin: Springer-Verlag)

\bibitem[{{Wicht}(2002)}]{Wicht02}
{Wicht}, J. 2002, Physics of the Earth and Planetary Interiors, 132, 281

\end{thebibliography}

\end{document}